\documentclass[11pt]{article}
\usepackage{moriond,epsfig}

\def\be{\begin{equation}}
\def\ee{\end{equation}}
\def\bea{\begin{eqnarray}}
\def\eea{\end{eqnarray}}

\begin{document}
\vspace*{4cm} \title{MASS PROFILES OF GALAXY CLUSTERS FROM THE
PROJECTED PHASE-SPACE DISTRIBUTION OF CLUSTER MEMBERS}

\author{ A. BIVIANO }

\address{INAF -- Osservatorio Astronomico di Trieste, 
via G.B. Tiepolo 11, \\ I-34131 Trieste, Italy}

\maketitle\abstracts{I review our current knowledge of the mass
distribution in clusters, as obtained from the analysis of the
projected phase-space distribution of cluster galaxies. I discuss the
methods of analysis, their relative advantages and disadvantages, and
their reliabilities. I summarize the most recent and important results
on the mass distributions of galaxy systems, from nearby to
medium-distant systems, and from groups to clusters. In particular I
consider how well different proposed models fit the observed cluster
mass distribution, and which are the relative distributions of
galaxies, baryons, and dark or total mass in clusters.  I also discuss
the current observational evidence for mass accretion onto galaxy
systems, coming mostly from the analysis of the velocity anisotropy
profiles.}

\section{Introduction}
The determination of cluster masses dates back to Zwicky's (1933,
1937) and Smith's (1936) preliminary estimates of the masses of the
Coma and the Virgo cluster. These early estimates were based on the
virial theorem with galaxies used as unbiased tracers of the cluster
potential (the {\em light traces mass} hyptohesis). These studies
marked the discovery of dark matter (see, e.g., Biviano 2000 for a
historical review of the subject). Only much later it was correctly
pointed out (by The \& White 1986 and Merritt 1987 among others) that
cluster masses were in fact poorly known, since relaxing the {\em
  light traces mass} assumption widens the range of allowed mass
models considerably.  However, the limited amount of redshift data
made it very difficult at that time, if not impossible, to constrain
the relative distributions of cluster mass and cluster light.

With the advent of multi-object spectroscopy, a large number of
redshifts for cluster galaxies has become available, in large part
thanks to the ESO Nearby Abell Cluster Survey (ENACS, Katgert et
al. 1996, 1998), and the Canadian Network for Observational Cosmology
(CNOC, Yee et al 1996; Ellingson et al. 1998).  Moreover, field galaxy
surveys have also contributed to increase the number of redshifts for
cluster galaxies, in particular, the Two Degree Field Galaxy Redshift
Survey (2dFGRS, e.g. Colless et al. 2001; De Propris et al. 2002) and
the Sloan Digital Sky Survey (SDSS, e.g. Adelman-McCarthy et al. 2006;
Miller et al. 2005).  More recent spectroscopic surveys of cluster
galaxies include the Las Campanas/Anglo­Australian Telescope Rich
Cluster Survey (Pimbblet et al. 2001), the Cluster and Infall Region
Nearby Survey (CAIRNS, Rines et al. 2003), the WIde Field Nearby
Galaxy-clusters Survey (WINGS, Fasano et al. 2006), and the ESO
Distant Cluster Survey (EDisCS, White et al. 2005). This large amount
of spectroscopic data now available for cluster galaxies provides the
observational material for the analysis of cluster mass
distributions. Knowledge of the mass distribution within clusters
(also in relation to the distributions of the different cluster
components) gives important clues about the way clusters of galaxies
and their components form and evolve (see, e.g., El Zant et al. 2004),
and about the nature of dark matter (see, e.g., Meneghetti et
al. 2001). In particular, the discovery of the universal 'NFW' mass
density profile of dark matter haloes by Navarro et al. (1996, 1997)
has stimulated many studies on the subject of mass profiles over the
last decade.

One way to determine the mass profile of a cluster is by the use of
its member galaxies as tracer of the gravitational potential. Another
very useful tracer is the intracluster (IC) diffuse gas component. In
both cases one uses the observed distribution of these tracers in
projected phase-space to infer the cluster mass distribution, very
often through the application of the spherically-symmetric,
collisionless Jeans equation (e.g. Binney \& Tremaine 1987). A more
direct method for the determination of a cluster mass profile is
through the analysis of the shapes and distribution of gravitationally
lensed galaxies in the cluster background. None of these methods is
without problems.  The lensing method suffers from the mass-sheet
degeneracy (e.g. Dye et al. 2001; Cypriano et al. 2004), projection
effects (e.g. Metzler et al. 2001; Wambsganss et al. 2005), and
low-$z$ inefficiency (e.g. Natarajan \& Kneib 1997). The main problem
with the use of the X-ray emitting IC gas as a tracer of the potential
is that X-ray observations generally sample only the inner cluster
regions (see, e.g., Pratt \& Arnaud 2002; but see also Neumann 2005).
Additional problems exist, such as incomplete thermalization and the
presence of bulk motions in the IC gas (e.g. Henriksen \& Tittley
2002) leading to the violation of the conditions of hydrostatic
equilibrium required for an unbiased determination of the cluster mass
(Bartelmann \& Steinmetz 1996; Kay et al. 2004; Rasia et al. 2004).

The problems that affect the cluster mass determinations obtained
using galaxies as tracers will be extensively discussed in
Sect.~2.1. Here I rather emphasize the main advantage of this method,
namely the fact that it allows sampling the mass profiles of galaxy
clusters out to large radii, well beyond the virialized region (see,
e.g., Reisenegger et al. 2000; Rines et al. 2002). An additional
advantage is that by using galaxies as tracers we can (or in fact,
{\em must}) learn about their orbital distribution within clusters
(see, e.g., van der Marel et al. 2000, vdM00 hereafter; Biviano \&
Katgert 2004), a useful piece of information for constraining models
of galaxy evolution in clusters. In the end, however, the main
rationale for using galaxies as tracers of the cluster potential is
that three is better than one!  Given that each method has its own,
difficult to correct for, biases, it is wise to determine cluster mass
profiles using several, independent methods.  In this paper, I review
recent results on cluster mass and mass-to-light ratio ($M/L$
hereafter) profiles as obtained from the analysis of the spatial and
velocity distributions of cluster member galaxies. I will also discuss
the evidence for mass accretion coming from these studies. The
interested reader may find complementary informations in my previous
reviews on these topics (Biviano 2002, 2005).

\section{Methods}
A cluster mass profile, $M(<r)$, can be determined from the projected
phase-space distribution of its member galaxies by means of the virial
theorem (see, e.g., Rines \& Diaferio 2006), the Jeans analysis (see,
e.g., Binney \& Tremaine 1987), and the 'caustic'
method recently introduced by Diaferio \& Geller (1997; see also
Diaferio 1999). When also galaxy {\em distance} estimates are
available, one can apply the least-action method (Peebles 1989) and
variants of it (see, e.g., Mohayee \& Tully 2005). Given the current
uncertainties on galaxy distances, this method is only applicable
to the very local Universe (typically, out to the Virgo cluster
distance).

Use of the virial theorem requires correction for the surface pressure
term, unless the radius at which the virial mass is estimated
effectively contains the whole system (The \& White 1986; Girardi et
al. 1998). The surface pressure term depends on the orbital
distribution of the population of tracers.

In the Jeans analysis the observable projected phase-space
distribution of galaxies is related to the cluster $M(<r)$, using the
Abel and Jeans equations (eqs. 4-55, 4-57, and 4-58 in Binney \&
Tremaine 1987, for the simplified case of a stable, non-rotating,
spherically symmetric system), through knowledge of the velocity
anisotropy profile, $\beta(r)$, which characterizes the orbits of
cluster galaxies.  The classical Abel inversion equation has recently
been extended by Mamon \& Bou\'e (in preparation) to the case of a
costant anisotropic velocity distribution.

In the caustic method, one infers $M(<r)$ of a given cluster from the
amplitude of the caustics in the space of line-of-sight velocities
vs. projected clustercentric distances.  Formally, the caustic
amplitude is related to the gravitational potential through
a function ${\cal F}$ of the potential itself, and of $\beta(r)$ (see
eqs. 9 and 10 in Diaferio 1999). Numerical simulations indicate ${\cal
F} \approx \rm{const}$, but only at radii larger than the cluster
virial radius, $r_{200}$ (see Fig.~2 in Diaferio 1999). Hence, for
$r<r_{200}$ the Caustic mass estimate is not very accurate. On the
other hand, since the method does not rely on the assumption of
dynamical equilibrium, it is a very powerful tool to constrain $M(<r)$
at large radii.

Note that with all these methods one samples the {\em total,} not the
{\em dark} mass of a cluster. This must be taken into consideration
when observational results are compared with results from numerical
simulations of collisionless dark matter haloes. The total matter
being the sum of the dark and baryonic matter, the dark matter
component can be estimated by subtracting the baryonic component from
the total matter. Most of the baryons are in the diffuse IC gas
component, hence X-ray data are needed in addition to galaxy data to
determine the dark matter distribution (see \L okas \& Mamon 2003;
Biviano \& Salucci 2006).

\subsection{Problems}
A fundamental problem of the Jeans analysis (and, to a lesser extent,
also of the virial and caustic analyses) is the 'mass--orbit'
degeneracy, i.e. the fact that the solution obtained for
$\mbox{$M(<r)$}$ is degenerate with respect to the solution obtained
for $\beta(r)$. In order to break this degeneracy, $\beta(r)$ must be
constrained independently from $M(<r)$. This can be achieved by the
comparison of the projected phase-space distribution of cluster
galaxies with distribution function models, where generally a constant
$\beta(r)$ is assumed (see, e.g., Kent \& Gunn 1982; vdM00; Mahdavi \&
Geller 2004), or via the analysis of the {\em shape} of the galaxy
velocity distribution, that contains the required information about
the orbital anisotropy of cluster galaxies (e.g. Merritt 1987; vdM00;
\L okas \& Mamon 2003). Alternatively, if several tracers of the
gravitational potential are available, the Jeans equation can be
solved for $M(<r)$ independently for each of the tracers, thus
restricting the range of possible solutions (Biviano \& Katgert 2004).

Another relevant problem in the virial and Jeans analysis occurs if
the cluster is not in steady state and dynamical equilibrium (this is
not a serious problem for the caustic method, see Rines et al.
2003). Since clusters grow by accretion of field galaxies (e.g. Moss
\& Dickens 1977; Biviano et al. 1997), they are not steady-state
systems.  Formally, inclusion of the time derivative in the Jeans
equation (eq. 4-29c in Binney \& Tremaine 1987) is then needed. On the
other hand, the fractional mass infall rate is estimated to be small
for nearby clusters (Ellingson et al. 2001). Moreover, most of the
mass is accreted in big, discrete clumps (Zabludoff \& Franx
1993). Hence, those clusters undergoing substantial mass accretion can
be identified through the presence of substructures in their galaxies
distribution, and these clusters (or their main subclusters) excluded
from the sample (vdM00; Biviano \& Girardi 2003; Katgert et al. 2004).
The problem gets tougher when one is dealing with small galaxy systems
(galaxy groups) most of which are probably still in a pre-virialized
collapse phase (e.g. Giuricin et al. 1988).

Interlopers are another potential problem in cluster mass estimates.
Interlopers are foreground/background galaxies in the cluster region
with velocities in the range of the velocity distribution of cluster
members. The methods for identifying interlopers have become
increasingly sophisticated over the years (e.g. Yahil \& Vidal 1977;
den Hartog \& Katgert 1996; Fadda et al. 1996; Carlberg et al. 1997a;
\L okas et al. 2006). These methods have been tested vs. numerical
simulations and shown to produce robust results (van Kampen \& Katgert
1997; Sanchis et al.  2004; \L okas et al.  2006; Biviano et al. 2006;
Wojtak et al. 2006). The different methods produce in general quite
consistent results for the cluster velocity dispersions (Girardi et
al. 1993).  It is nevertheless advisable to use robust estimators of
the moments of the velocity distribution, both for the velocity
dispersion (such as the biweight, see Beers et al. 1990) and even more
so for the kurtosis (often used to constrain $\beta$, see \L okas \&
Mamon 2003), which can be effectively replaced by the Gauss-Hermite
moments (see vdM00).

As the number of available redshifts for cluster galaxies increases,
the interloper selection procedure becomes more robust (e.g. Biviano
et al. 2006).  Since in practice spectroscopic samples of several
hundreds member galaxies per cluster are rare, quite often a
'composite' cluster is built by stacking together the data for several
clusters (see, e.g., Carlberg et al. 1997a; vdM00; Katgert et
al. 2004). The procedure is justified if clusters form a homologous
family, as suggested by the existence of a fundamental plane of
cluster properties (Schaeffer et al. 1993; Adami et al. 1998), and if
the mass range of the stacked clusters is not too large, since the
concentrations of dark matter haloes are expected to depend (albeit
only mildly) on the halo masses (e.g. Dolag et al. 2004).

In the Jeans analysis it is generally assumed that clusters are
spherical and do not rotate. The composite cluster is spherically
symmetric by construction, and deviation from spherical symmetry is
unlikely to be a major problem for individual clusters either (see
vdM00; Sanchis et al. 2004). While evidence for cluster rotation has
been claimed for a couple of clusters (see, e.g., Dupke \& Bregman
2001), the energy content in the rotational component is marginal.

Dynamical friction and galaxy mergers could in principle invalidate
the use of the {\em collisionless} Jeans equation. However, the
cluster velocity dispersion is too high for galaxy mergers to take
place, and the dynamical friction timescale is too long for most
cluster galaxies, except for very massive galaxies (e.g. Biviano et
al. 1992). These can be removed from the sample (Katgert et al. 2004)
before going through the Jeans analysis. The situation is probably
different in low-velocity dispersion systems, like groups, where the
galaxy merging process may be rather efficient (see, e.g., Conselice
2006).

\subsection{Reliability}
The reliability of the different estimators of mass profiles can be
assessed via numerical simulations, and via direct comparison of the
results obtained with the different estimators. Comparison with
numerical simulations indicate that the Jeans and caustic methods
produce reliable mass profile estimates (Diaferio 1999; Sanchis et
al. 2004; \L okas et al. 2006). Comparison of the mass profiles
obtained for 72 nearby clusters using either the caustic or virial
methods show consistency in 2/3 of the cases (Rines \& Diaferio 2006).
The mass profile of an ensemble of 43 nearby clusters determined via
the Jeans analysis assuming isotropic orbits is found to be consistent
with the one determined via the caustic analysis (Biviano \& Girardi
2003). Viceversa, the mass profile of an ensemble of 8 nearby clusters
determined via the caustic analysis is found to be consistent with the
one determined via the Jeans analysis if the orbits are close to
isotropic (Rines et al. 2003).

Very little is known so far about the consistency of mass profiles
obtained using either cluster galaxies or the IC gas as tracers.  The
agreement between the X-ray- and caustic-determined masses is not very
good for the sample of three medium-$z$ clusters analysed by Diaferio
et al. (2005). A better agreement is found by Benatov et al. (2006)
for two of the same clusters analysed by Diaferio et al. (2005), as
well as for another three nearby clusters, using the Jeans rather than
the caustic method.  In this case, different orbital anisotropies are
inferred for the different clusters (although the significance of
these differences is rather low).  The lensing mass profiles obtained
for the three clusters analysed by Diaferio et al. (2005) are in
reasonable agreement with the caustic mass profiles, and the lensing
mass profile obtained for the A2218 cluster analysed by Natarajan \&
Kneib (1996) is consistent with the Jeans mass profile, is one drops
the isotropic assumption.

\section{Mass profiles}
In a series of papers, Geller et al. (1999) and Rines et al. (2000,
2001, 2003, 2004) applied the caustic technique to 9 nearby clusters
of the CAIRNS survey (including the Coma cluster). They found that the
mass density profile $\rho(r)$ decreases like $\rho(r) \propto r^{-1}$
near the center, and the best-fit asymptotic slope of the mass density
profile at large $r$ is either $-3$ (NFW) or $-4$ (Hernquist
1990). When the NFW profile is adopted, the best-fit values of the
concentration parameter $c$\footnote{Throughout this paper I adopt the
definition $c \equiv r_{200}/r_s$, where $r_s$ is the scale radius
parameter in the NFW profile, see Navarro et al. (1997).} vary between
5 and 17. Using the mass profiles obtained from the caustic technique
in the Jeans analysis, $\beta \approx 0$ is obtained.  These results
have recently been confirmed by a new caustic analysis of a sample of
72 clusters extracted from the SDSS (Rines \& Diaferio 2006).

Biviano \& Girardi (2003) analysed a composite cluster of 1345 member
galaxies, obtained from stacking together 43 nearby clusters from the
2dFGRS. They performed a joint Jeans and caustic analysis, using the
former method to determine the mass profile in the virialized region
($\leq r_{200}$), and the latter to extend $M(<r)$ to $2 \,
r_{200}$. I.e. they used each of the two methods in the cluster region
where it is expected to perform best. A $c \simeq 6$ NFW profile was
found to fit the data over the whole radial range explored.  Cored
profiles are also acceptable but only if the size of the core radius
is small, $< 0.1 \, r_{200}$.

\begin{figure}
\center{\psfig{figure=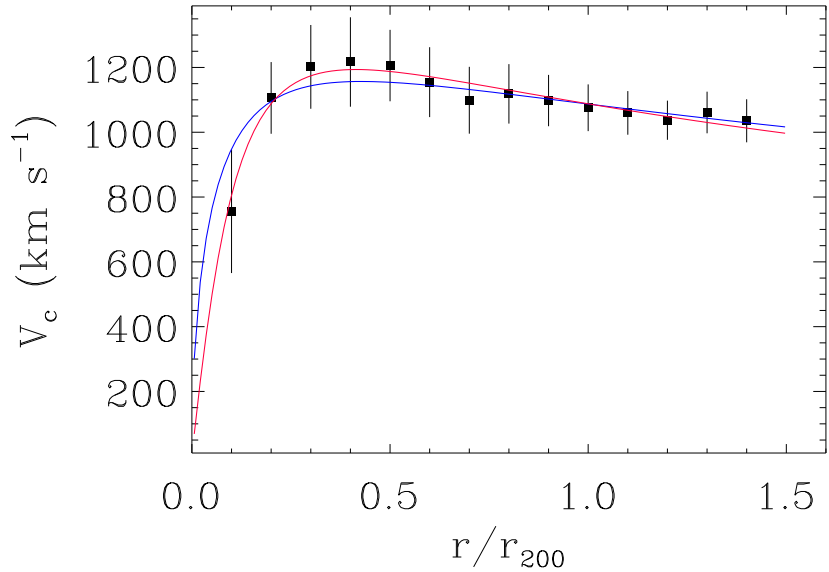,height=3.5in}
\caption{The circular velocity profile, $V_c \equiv (GM/r)^{0.5}$, of
the dark matter component in a composite of 59 ENACS clusters.
1-$\sigma$ error bars are shown.  The two lines show the best-fit
Burkert (1995) and NFW models (red and blue line, respectively -- the
NFW model fit predicts slightly higher circular velocity at small
radii than the Burkert (1995) fit).  Adapted from Biviano \& Salucci
(2006).}}
\end{figure}

\begin{figure}
\center{\psfig{figure=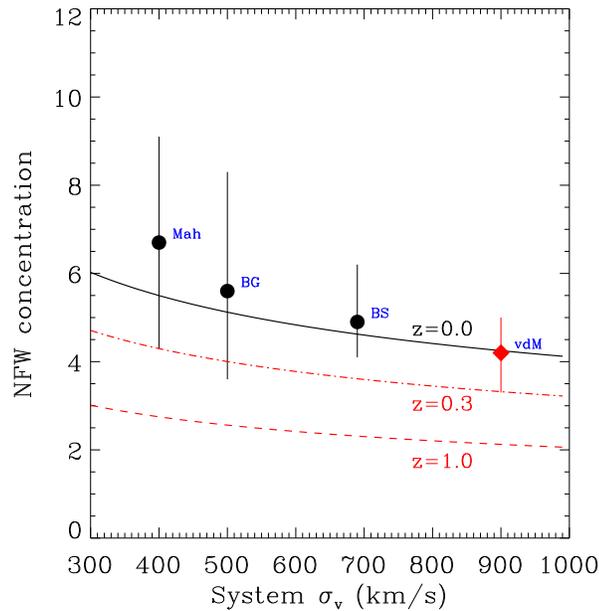,height=3.5in}
\caption{The best-fit concentration parameter of the NFW profile as a
function of the average velocity dispersion for five different
data-sets of galaxy systems (Mah=Mahdavi et al. 1999, BG=Biviano \&
Girardi 2003, BS=Biviano \& Salucci 2006, vdM=vdM00). 1-$\sigma$ error
bars are shown. Dots are for nearby galaxy systems, the diamond is for
the CNOC sample, at a median $z=0.3$. Predictions from Dolag et
al. (2004) are shown as lines for three different redshifts.}}
\end{figure}

Katgert et al. (2004) analysed a composite of 1129 galaxies from 59
nearby clusters from the ENACS. $M(<r)$ was determined in a fully
non-parametric way.  Only early-type galaxies (ellipticals and S0s)
were used in the analysis. Their velocity distribution was showed to
imply nearly isotropic orbits, hence $\beta \equiv 0$ was adopted in
the Jeans analysis. The resulting mass density profile has a slope of
$-2.4 \pm 0.4$ at $r=r_{200}$, fully consistent with an asymptotic
slope of $-3$ at larger radii. Both a $c=4 \pm 2$ NFW profile, and a
Burkert (1995) profile with a rather small core radius ($\leq 0.1 \,
r_{200}$) are acceptable. The $M(<r)$ solution obtained by Katgert et
al. (2004) was confirmed by Biviano \& Katgert (2004) by using another
tracer of the gravitational potential, early-type {\em spirals}
(Sa--Sb) in lieu of early-type galaxies.

\L okas \& Mamon (2003) and \L okas et al. (2006) determined the mass
distribution of 7 nearby clusters (including Coma) using the Jeans
method, and fitting simultaneously the velocity dispersion and
kurtosis profile, in order to break the mass-anisotropy degeneracy
(see Sect.~2.1).  Their cluster mass profiles are generally well
described by the NFW model, with $c$ ranging from 2 to 11, with a
median of 7.

All of these works, except for the analysis of the Coma cluster by \L
okas \& Mamon (2003), actually constrained the {\em total,} not the
{\em dark} mass profile. Recently, Biviano \& Salucci (2006) have
re-analysed the ENACS data-set to split the total matter profile into
its dark and baryonic components. They find that the dark matter
profile is consistent with both a Burkert (1995) and a NFW
model. Their best-fitting models (see Figure~1) are slightly more
concentrated than those fitting the total mass profile. They hardly
change whether one considers the whole dark matter, or only its {\em
diffuse} component, obtained by subtracting the contribution from
individual dark matter subhaloes.

Similar results are obtained for the $M(<r)$ of higher-$z$ clusters.
Following up the previous analyses of Carlberg et al. (1997a, 1997b),
vdM00 stacked together 16 clusters at $z=0.17$--0.55 from the CNOC
survey, to build a composite cluster with $\simeq 990$ galaxies.  From
the Jeans analysis they then determined a mass profile which is fully
consistent with a $c \simeq 4$ NFW model.

At the low-mass end, results are still controversial.  Mahdavi et
al. (1999) used 588 galaxies in 20 groups to conclude that their mass
density profile is consistent with a Hernquist (1990) model.  Mahdavi
\& Geller (2004) determined the average mass density profile of groups
from the RASSCALS sample (893 galaxies in 41 nearby groups, see
Mahdavi et al. 2000). They find it to be well described by a single
power-law model over a wide radial range (0--$2 \, r_{200}$). Finally,
Carlberg et al. (2001) analysed $\sim 800$ galaxies in $\sim 200$
groups from the CNOC2 survey, at redshifts $z=0.1$--0.55. They
determined a very shallow mass density profile, with a central core.

Discrepant results on the mass profiles of groups probably origin from
the fact that not all groups are dynamically virialized systems
(Giuricin et al. 1988; Diaferio et al. 1993; Mamon 1995; Mahdavi et
al. 1999). Alternatively, since the merger process is expected to be
efficient in some groups (Mamon 1995), the use of the collisionless
Jeans equation may not be justified (see Sect.~2.1).  Mamon et al. (in
preparation) are currently trying to determine the average mass
profile of the GEMS groups (Osmond \& Ponman 2004), for which both
X-ray and optical data are available. The comparison of optical and
X-ray properties should help constraining the groups dynamical status
(see, e.g. Brough et al. 2006). Preliminary results seem to indicate
that the virialized groups have mass profiles consistent with the NFW
model, with a higher concentration parameter than clusters.

By assembling several results from the literature, I show in Figure~2
how the best-fit concentration parameter $c$ of the NFW model changes
with the velocity dispersion of the systems considered. The trend is
consistent with expectations from cosmological numerical simulations.

\section{Mass accretion}
The hierarchical model for the formation of cosmological structures
has been so successful over the years that it has now become a
paradigm. A way to test the paradigm is to look for evidence of the
hierarchical build-up of cosmological structures. Direct evidence for
the infall of field galaxies into galaxy systems has been recently
provided by Ceccarelli et al. (2005), from the analysis of the pattern
of peculiar velocities around nearby groups. In other galaxy systems,
for which galaxy distances are generally not available, evidences for
mass accretion can be obtained from the analysis of their mass
profiles. In particular, using cluster mass profiles determined via
the caustic method, Rines \& Diaferio (2006) have recently estimated
that the mass within the turnaround radius of a cluster is about twice
the mass within its virial radius, on average. Hence nearby clusters
are still forming and will eventually double their virial mass in the
future.

Another evidence for mass accretion comes from the Jeans analysis.
Most studies find evidences that the velocity anisotropies for cluster
and group galaxies are not strong, $\beta \approx 0$ (Carlberg et al.
1997b; Mahdavi et al.  1999; vdM00; \L okas \& Mamon 2003; Rines et
al. 2003; Katgert et al. 2004; \L okas et al. 2006), particularly so
when only early-type galaxies are considered. The situation is
different when late-type galaxies are examined. The suggestion that
late-type galaxies may be an infalling population dates back to Moss
\& Dickens' (1977) discovery that this population has a wider velocity
distribution than the early-type population. Through the modelling of
the projected phase-space distribution of blue galaxies in the CNOC
clusters, Carlberg et al. (1997b) suggested that these galaxies are
actually in equilibrium with the cluster potential, but on slightly
radial orbits ($\beta \sim 0.5$). The same conclusion was reached by
Mahdavi et al. (1999) for nearby groups, and by Biviano (2002) and
Biviano \& Katgert (2004) for nearby clusters.

Given the mass profile determined by Katgert et al. (2004), Biviano \&
Katgert (2004) inverted the Jeans equation (following Binney \& Mamon
1982 and Solanes \& Salvador-Sol\'e 1990) and determined the velocity
anisotropy profiles of different classes of galaxy types. They found
that the isotropic solution is acceptable for ellipticals and S0s, for
early-type spirals (Sa--Sb), but not for late-type spirals (Sbc-type
and later).  These galaxies are characterized by $\beta \approx 0$
only near the center, and have increasingly radial orbits
outside\footnote{A similar, albeit preliminary, analysis performed on
a larger cluster sample extracted from the SDSS, confirm this result
with greater statistical accuracy.}, a profile remarkably similar to
those obtained for dark matter particles in numerical simulations
(e.g. Ghigna et al. 1998; Diaferio 1999). By analogy, it is tempting
to suggest that late-type spirals, just like dark matter particles in
numerical simulations, still retain memory of the process of (mostly
radial) gravitational infall along the filaments connecting to the
cluster, and are therefore recent arrivals into the cluster (which
would also explain why their internal properties have not been
affected by the cluster potential, see the discussion in Biviano \&
Katgert 2004).

The projected phase-space distributions of early- and late-type
galaxies in the nearby ENACS clusters are remarkably similar to those
of, respectively, red and blue galaxies in the medium-$z$ CNOC
clusters. Unsurprisingly then, also their orbital distributions are
similar, nearly isotropic that of early-type ENACS and red CNOC
galaxies, and mildly radial that of late-type ENACS and blue CNOC
galaxies. The only remarkable difference is the substantially higher
fraction of blue galaxies in the CNOC clusters, as compared to the
fraction of late-type galaxies in the ENACS clusters. If
blue/late-type galaxies are newcomers from the field in the cluster
potential, as suggested by their radial orbits, we infer that the
accretion rate of field galaxies into clusters was higher in the past
(Ellingson et al. 2001). In order to reduce the late-type cluster
galaxy fraction from $z \sim 0.3$ to $z \sim 0$, blue, late-type
cluster galaxies must either transform and/or dim with time. If they
transform, changing their type and colour, they need at the same time
change their orbits, from radial to nearly isotropic. This is a
powerful constraint for models of galaxy evolution in clusters, since
not all transformation mechanisms are able to affect galaxy properties
as well as their orbits.

\section{$M/L$ profiles}
Most of the studies addressing the determination of cluster $M(<r)$s,
also address the determination of cluster $M/L$s (in particular:
Biviano \& Girardi 2003; Katgert et al. 2004; Rines et al. 2004).  We
therefore refer the reader to Section~3 for details on the data
samples used. The results obtained by the different authors show a
great degree of consistency, modulo the selection in type or colour of
the cluster galaxy population used to build the luminosity (or number)
density profile. In fact, it is well known that galaxies of different
types have different luminosity and number density profiles (Dressler
1980; see also Biviano et al. 2002 and references therein).

For nearby clusters it is generally found that the $M/L$ profile is
rather flat out to $r_{200}$, and decreases beyond that, by a factor
$\times 2$ out to the turnaround radius. The decreasing trend is due
to the increasing contribution of late-type galaxies in the external
regions. Approaching the cluster centers (at radii $\leq 0.2 \,
r_{200}$) there is a drop of $M/L$ mainly caused by the presence of a
brightest cluster galaxy (BCG) at the bottom of the cluster potential
well. When red (or early-type) galaxies are selected, or when the
selection of cluster galaxies is made in a very red photometric band
(e.g. the $K$ band), {\em and} the cluster BCG contribution is
removed, the light of cluster galaxies is found to trace the
mass. This is good news for the determination of masses using the
virial theorem (see Sect.~1), which is therefore expected to produce
unbiased results when the red sequence cluster galaxies are used as
tracers (as confirmed by the analysis of clusters extracted from a
cosmological simulation, see Biviano et al. 2006). 

The $M/L$ profile of more distant clusters is not so well constrained
as that of nearby clusters. Like the $M/L$ profile of nearby clusters,
it is consistent with being $\approx$ constant out to the virial
radius, if only red galaxies are selected (vdM00).

At lower masses, results on the $M/L$ profile are
controversial. Popesso et al. (2006) have recently shown that the
central slope of the number density profiles of galaxy systems
flattens with decreasing system mass.  Mass density profiles are
instead thought to be more concentrated for less massive haloes
(e.g. Navarro et al. 1997; Dolag et al. 2004). Hence galaxy groups
should have decreasing $M/L$ profiles with radius. Instead, group
$M/L$ profiles are found to be flat (Mahdavi et al. 1999), if not even
increasing (Carlberg et al. 2001). The discrepant results for the
group $M/L$ profiles are a consequence of the poor knowledge of the
group mass profiles (see Sect.~3).

Widening the scope of the $M/L$ analyses, Biviano \& Salucci (2006)
have recently determined the baryonic-to-total mass fraction as a
function of radius for a sample of 59 nearby clusters from the ENACS
data-base. They find that the total baryonic mass is less concentrated
than the total mass, except in the central region.  The excess
baryonic mass near the center is due to the BCG, while the overall
trend at radii $\geq 0.25 \, r_{200}$ is due to the IC gas that
dominates the cluster baryon budget.  The IC gas-to-total mass
fraction increases with radius as $r^{0.4}$.

\section{Summary and perspectives}
Useful constraints on the mass distribution within clusters have been
obtained in recent years from the analysis of the projected
phase-space distribution of cluster member galaxies. The constraints
near the center ($\rho(r) \propto r^{\xi}$ with $-2 \leq \xi \leq 0$)
are not strong because of an intrinsic finite resolution (the size of
the BCG). Both cored and cuspy profiles are allowed.  Stronger
constraints are obtained at large radii ($\rho(r) \propto r^{\zeta}$
with $-4 \leq \zeta \leq -3$), and they are stronger than those
obtained with any other methods.  The isothermal sphere is ruled out,
while the NFW, Burkert (1995) and Hernquist (1990) models are
acceptable. If the mass density profile has a core, this has to be
small, i.e. the radius where the mass density drops to half its
central value is $\leq 0.1 \, r_{200}$. By comparison with the results
of numerical simulations of Meneghetti et al. (2001), Katgert et
al. (2004) have used this result to set an upper limit to the
scattering cross-section of dark matter particles. If, on the other
hand, NFW is taken to be the correct model, then its concentration
parameter is found to scale with the system mass as expected from
cosmological numerical simulations (see Figure~2).

Old, passively evolving cluster galaxies have a nearly isotropic
orbital distribution. On the other hand, the radial orbital
distribution of late-type galaxies suggests they are recent newcomers
into the cluster environment, which is therefore still accreting mass
from the surrounding field, but at a lower rate than they used to
at $z \sim 0.3$.

Overall, galaxy luminosity has a more concentrated distribution than
cluster mass, but mass is traced quite well by galaxy light within the
cluster virial radius, when only red, early-type galaxies (excluding
the BCG) are considered. Blue, late-type galaxies, as well as the IC
gas, have a more extended distribution than the mass.

The average mass profiles of nearby clusters and medium-distant
clusters are similar, insofar as relatively relaxed clusters are
selected. Hence, little or no evolution of the cluster dynamics from
$z \sim 0$ to $z \sim 0.5$ is implied (see also Girardi \& Mezzetti
2001), but this may simply reflect an observational bias, since
clusters undergoing mergers are not eligible for the classical
dynamical analyses. When comparing the $M/L$ and velocity anisotropy
profiles of nearby and medium-distant clusters, one needs to take into
account that the fraction of late-type galaxies increases with $z$
(e.g. Dressler et al. 1997; Fasano et al. 2000). When the comparisons
are made {\em separately} for early- and late-type galaxies, there is
no evidence for significant evolution of either the $M/L$ or the
velocity anisotropy profiles\footnote{Benatov et al.'s (2006)
suggestion that the velocity anisotropy profile of clusters evolves
with $z$, finds its natural explanation in the evolving fractions of
early- and late-type galaxies, the former on isotropic and the latter
on radial orbits.}.

All the above results must be considered valid for clusters {\em on
average,} since projection effects and cosmic variance may lead to
rather different results for different individual clusters (e.g. Rines
et al. 2003; Sanchis et al. 2004).

Progress on these topics is to be expected from the use of the SDSS
data-set (e.g. Miller et al. 2005), which will allow to reduce the
current uncertainties on the dynamics of nearby clusters.  Much is
still to be learned about the dynamics of lower-mass galaxy systems
(groups). It is important to determine the mass profile of galaxy
groups, intermediate in size between clusters and galaxies, because
galaxy clusters have mass profiles that are reasonably well described
by the NFW profile, but galaxies may not (de Blok \& Bosma 2002; de
Blok et al. 2003; Borriello et al. 2003; Gentile et al. 2004).  A
promising data-set in this respect is the GEMS groups sample of Osmond
\& Ponman (2004), for which also X-ray temperatures are
available. These can provide robust scaling parameters for the
build-up of a composite group sample. Finally, as more and more data
become available for $z>0.5$ clusters (e.g. Demarco et al. 2005;
Girardi et al. 2005; White et al. 2005), it will perhaps be possible
to determine the evolution of cluster mass, $M/L$, and velocity
anisotropy profiles, although the young dynamical age of high-$z$
clusters is an additional problem we will have to deal with.

\section*{Acknowledgments}
I dedicate this paper to the memory of my dear friend Daniel Gerbal.
I wish to thank Sophie Maurogordato and Laurence Tresse for organizing
such an exciting and enjoyable meeting, and the Program and Scientific
Advisory Committees for inviting me to give this review, as well as
for providing partial financial support. This work has been partially
supported by funds from the INAF Progetto di Ricerca di Interesse
Nazionale: ``The WINGS Project and Follow-Ups''.

\end{document}